\documentclass[showpacs,showkeys,aps,prc,amsmath,amssymb,twoside]{revtex4}

\usepackage{graphicx}
\usepackage{dcolumn}
\usepackage{bm}

\begin{document}


\title{The hypernuclear Auger effect within the density dependent relativistic
  hadron field theory}

\author{Christoph Keil}
\author{Horst Lenske}%
\affiliation{%
  Institut f\"ur Theoretische Physik, University Gie\ss en\\
  Heinrich-Buff-Ring 16\\D-35392 Gie\ss en
  }%

\date{\today}

\begin{abstract}
The hypernuclear Auger effect, given by the de-excitation of a
$\Lambda$ hypernucleus by means of the transition of a $\Lambda$
hyperon from an initial to a lower lying final single particle
state in conjunction with neutron emission from the core nucleus
is studied in relativistic DDRH field theory. Baryonic
interactions are obtained from the Bonn NN potential by
Dirac-Brueckner theory and theoretically derived scaling laws for
the meson-hyperon vertices. The model is applied to the
$^{209}_\Lambda$Pb hypernucleus. $\Lambda$ and nucleon bound
states as well as scattering states are calculated
self-consistently in mean-field approximation. The Auger spectra
of the emitted neutrons are of complex structure due to a huge
combinatorial number of possible hypernuclear transitions. The
sensitivity of the neutron Auger spectra to changes in the
$\Lambda\sigma$ and $\Lambda\omega$ vertices is investigated. The
theoretical results show that experimental applications of the
hypernuclear Auger effect will require special efforts, e.g. by
tagging on the energy of the initially created hyperon.
\end{abstract}

\pacs{21.80.+a}
\keywords{hypernuclei, spectroscopy, Auger}
\maketitle
\section{\label{sec:intro}Introduction}
In order to obtain a more profound understanding of the
interactions between baryons with and without strangeness the
major goal of hypernuclear physics is to explore the spectral
properties of $\Lambda$ hyperons in a nucleus. The experimentally
well confirmed existence of $\Lambda$ single particle states
\cite{Hotchi:2001rx,nagae-hyp2000,may-c13,Hasegawa:fj,Ajimura:1994jb,Pile:cf}
indicates the importance of static mean-field interactions. While
the overall features are reasonably well reproduced by
non-relativistic \cite{Hjorth-Jensen:1996yj,Vidana:1998ed} and
relativistic approaches
\cite{Rufa:uf,Rufa:zm,Glendenning:1992du} open
questions exist on finer details of hypernuclear spectra. In light
nuclei, $A \leq 16$, the smallness of the spin-orbit splitting and
other fine structures of $\Lambda$ single particle spectra are
well described by multi-configuration shell model calculations
with empirical matrix elements \cite{Millener:2000th}. These
calculations emphasize the importance of residual interactions and
configuration mixing. Theoretical studies of dynamical
self-energies in $\Lambda$ nuclei, predicting rather strong
contributions from the coupling of the $\Lambda$ to excitations of
the nuclear core, point in a similar direction
\cite{Hjorth-Jensen:1996yj,Vidana:1998ed}. This, however, is not
fully complying with recent observations of well separated and
relatively sharp $\Lambda$ single particle structures in the
medium-mass $^{51}_\Lambda$V and $^{89}_\Lambda$Y hypernuclei
\cite{Hotchi:2001rx,nagae-hyp2000}. The measured spectra are
compatible with a much stronger spin-orbit strength than expected
from the low mass region, amounting to spin-orbit interaction
energies of 1 to 2 MeV, but the data do not indicate a strong
damping of the $\Lambda$ single particle states.

The conclusion to be drawn from these - partly conflicting -
findings is that there is a clear necessity for more precise
measurements and an enlarged body of data. For more decisive
results on $\Lambda$-nucleus interactions medium and heavy mass
nuclei can be expected to be much better suited. In light nuclei
threshold effects from the weak binding of $\Lambda$ states and
finite size effects contribute strongly to the observed spin-orbit
splitting, thus not giving direct access to the wanted information
on the genuine $\Lambda N$ interaction strengths
\cite{Keil:2001rm}. The calculations in ref.~\cite{Keil:2000hk}
indicate that these effects decrease rapidly beyond the silicon
mass region. A standard tool for spectral investigations is
$\gamma$-spectroscopy, detecting the photons from the hypernuclear
transitions of the $\Lambda$ particle after it was created in an
excited state e.g. in a $A(\pi, K)_{\Lambda}A$ reaction. In a
recent $(K^-,\pi^-)$ experiment on $^{13}$C  at BNL/AGS progress
has been made by achieving a much better energy resolution than
before \cite{Kohri:2002nc}. These measurements, by the way,
confirm the small $\Lambda$ spin-orbit splitting in light nuclei.
By technical reasons $\gamma$-spectroscopy is not applicable in
medium and heavy mass nuclei.

A promising alternative for spectroscopy in medium and heavy mass
hypernuclei is the observation of Auger neutrons, emitted during
the de-excitation of the hypernucleus after the initial creation
of a $\Lambda$ in an excited state. The hypernuclear Auger process
was discussed first by Likar {\it et al.} \cite{Likar:1986jj}.
More recently, the idea has been revived and worked out in much
more detail for an experiment proposal at JLAB \cite{Margaryan}.
\begin{figure}[htbp]
  \begin{center}
    \includegraphics[width=.2\linewidth]{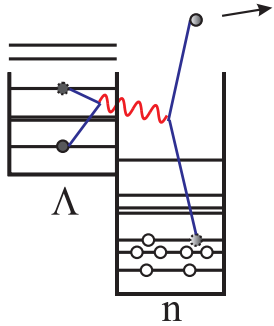}
    \caption{Schematic picture of the hypernuclear Auger effect: A
      $\Lambda$ hyperon is de-excited into a lower single particle
      level,thereby
      transferring energy and momentum to a valence neutron which is
      emitted through this process.}
    \label{fig:auger_scheme}
  \end{center}
\end{figure}
The hypernuclear Auger effect is the direct de-excitation of an
excited $\Lambda$ single-particle state in a hypernucleus by the
emission of a neutron. The process is analogous to the well known
atomic Auger effect where the de-excitation of an electron kicks
out another less bound one. In the case of the hypernuclear Auger
effect the energy spectrum of the emitted neutrons reflects the
$\Lambda$ singe-particle level structure, but folded with the
neutron single-particle spectrum. The mechanism of the
hypernuclear Auger effect is schematically shown in
figure~\ref{fig:auger_scheme}. 
In order to occur the separation energy of at least the $\Lambda$ 1s-orbit
must be larger than the separation energy of the valence neutron. Therefore,
because of the much weaker $\Lambda$ binding, the hypernuclear Auger effect
appears with a strength sufficient for measurements only in intermediate and
heavy mass nuclei where neutron and $\Lambda$ separation energies of
comparable magnitude are available.
Already from this simplified
picture it is clear that a single transition of the $\Lambda$ will
yield a wealth of peaks in the neutron's energy distribution,
requiring a very careful analysis. For the proposed JLAB
experiment \cite{Margaryan} a neutron energy resolution of better
than 50 keV is envisaged. As nuclei to be studied in the
experiment $Pb$ and $U$ are proposed.
In this mass region it can yield
valuable information about the $\Lambda$ single particle spectra,
possibly even resolving finer details as the spin-orbit splitting.
One purpose of this paper is to investigate in a realistic model
predictions for spectral distributions which may be used to
estimate constraints on future measurements.

In this paper we report on our calculations of the hypernuclear
Auger spectra within the density dependent relativistic hadron
field theory (DDRH). In section~\ref{xec:aug} we describe the
theoretical description of the hypernuclear Auger effect in the
relativistic DDRH approach with density dependent meson-baryon
vertices. We shortly sketch the main features of DDRH and the
numerical realization. In section~\ref{sec:results} results are
presented and possibilities to extract information from the
extremely complex spectra are discussed. In
section~\ref{sec:conclusions} we summarize, discuss our results
and draw some conclusions.

\section{\label{xec:aug}The Hypernuclear Auger Effect}
\subsection{\label{ssec:formalism}Formalism}

The emission of a neutron by the $\Lambda$ de-excitation of the
hypernucleus is described as a decay of an initial $\Lambda$
single particle state into a 2 particle-1 hole configuration where
the $\Lambda$ is coupled to a neutron particle-hole core
excitation with an energy above the particle emission threshold.
The nucleon and hyperon single particle states are obtained in
relativistic mean-field (RMF) approximation. Residual interactions
among the neutron particle-hole configurations are neglected, i.e.
an uncorrelated quasiparticle description is used. 

For the transition operator $V$ the one-boson-exchange
parametrization of the full Dirac-Brueckner (DB) G-matrix with
density dependent vertex functionals $\Gamma(\hat\rho)$
\cite{Fuchs:1995as} is used. It is the same interaction as applied
in the structure calculation. Since the $\Lambda$ hyperon is
electrically neutral and an iso-scalar particle only the $\sigma$
and the $\omega$ mesons contribute:
\begin{equation}
  \label{eq:pot}
  V = \Gamma_{\sigma\Lambda}(\hat\rho_\Lambda)
  \frac{1}{q^2-m_\sigma^2}
  \Gamma_{\sigma N}(\hat\rho_N)
  -\Gamma_{\omega\Lambda}(\hat\rho_\Lambda)
  \frac{1}{q^2-m_\omega^2}
  \Gamma_{\omega N}(\hat\rho_N)
\end{equation}

Taking $\left| 0 \right>$ to be our many-body 
ground state, assumed here as the
$0^+$ ground state of a spherical nucleus, the initial state is in
second quantization formulation given by $a^\dag_{\Lambda_\alpha}
\left| 0 \right>$, where $a^\dag_{\Lambda_\alpha}$ is the creation
operator for a $\Lambda$ state with the set of quantum numbers
$\Lambda_\alpha$. In the final state the hyperon is attached to a
particle-hole excited nuclear core. The excess energy and momentum
is carried away by the emitted neutron occupying an unbound single
particle continuum state in the nuclear mean-field potential:
\begin{equation}
  \label{eq:final-state}
  \left[ a^\dag_{\Lambda_\beta}\otimes
    A^\dag_{n_\beta}(j_{n_\beta}j_{n^{-1}_\beta})\right]
  _{j_\beta m_\beta} \left| 0 \right> =
  \sum_{m_{\Lambda_\beta}M_{n_\beta}}\left< j_{\Lambda_\beta}m_{\Lambda_\beta}
    J_{n_\beta}M_{n_\beta} \right|\left. j_\beta m_\beta \right>
  a^\dag_{\Lambda_\beta}
  A^\dag_{n_\beta}(j_{n_\beta}j_{n^{-1}_\beta}) \left| 0 \right>
\end{equation}
$A^\dag_{n_\beta}$ is the particle-hole excitation operator with
angular momentum $J_{n_\beta}, M_{n_\beta}$ defined through:
\begin{equation}
  \label{eq:ph-excitation}
  A^\dag_n(j,j')
  = \sum_{m,m'}\left<jmj'm'\right|\left. J_nM_n\right>
  a^\dag_{jm}\tilde{a}_{j'm'},
\end{equation}
where $\tilde{a} = (-)^{j+m}a_{j,-m}$ denotes a hole creation
operator. In eq.\ref{eq:final-state} the $a^\dag_{\Lambda_\beta}$
and $A^\dag_{n_\beta}$ are coupled to total angular momentum
$j_\beta$, $m_\beta$.

The differential widths $d\Gamma_{j_{\Lambda_\alpha}}$ describing the
decay of an initial $\Lambda_\alpha$ state is determined by the
transition matrix elements of the $\Lambda N$ interaction $V$:
\begin{equation}
  \label{eq:differential-width1}
  d\Gamma_{j_{\Lambda_\alpha}} = \frac{1}{32\pi^2}
  \frac{1}{2j_{\Lambda_\alpha}+1}
  \sum_{\left\{\gamma_1\right\}} \left|
    \left<0\right| \left[a_{\Lambda_\beta}\otimes
      A_{n_\beta}(j_{n_\beta},j_{n^{-1}_\beta})
    \right]_{j_\beta m_\beta} V a^\dag_{\Lambda_\alpha}
    \left| 0 \right>
  \right|^2 \frac{|\vec k|}{M^2}d\Omega,
\end{equation}
where $\quad \left\{\gamma_1\right\} = \left\{
m_{\Lambda_\alpha},m_\beta, j_\beta, j_{\Lambda_\beta},
  j_{n_\beta}, j_{n^{-1}_\beta} \right\}$ indicates the incoherent
summation over degenerate initial and final sub-states, including
the appropriate phase space factors \cite{pdg} due to the neutron
emerging with momentum $\vec{k}$. $M$ is the mass of the initial
hypernucleus. The orthogonality and completeness relations of the
Clebsch-Gordan coefficients \cite{glaudemans} allow to convert
this expression into the equivalent form of an incoherent sum over
matrix elements of uncoupled states:
\begin{equation}
 \label{eq:differential-width2}
  d\Gamma_{j_{\Lambda_\alpha}} = \frac{1}{32\pi^2}
  \frac{1}{2j_{\Lambda_\alpha}+1}
  \sum_{\left\{\gamma_2\right\}} \left|
    \left<0\right|
    a_{\Lambda_\beta}a_{n_\beta}\tilde{a}^\dag_{n^{-1}_\beta}
    V a^\dag_{\Lambda_\alpha}
    \left| 0 \right>
  \right|^2 \frac{|\vec k|}{M^2}d\Omega \quad ,
\end{equation}
where $ \left\{\gamma_2\right\} = \left\{ m_{\Lambda_\alpha},
m_{\Lambda_\beta}, m_{n_\beta},m_{n^{-1}_\beta},
j_{\Lambda_\beta},j_{n_\beta}, j_{n^{-1}_\beta} \right\}$. From
equations~(\ref{eq:differential-width1}) and
(\ref{eq:differential-width2}) it is
seen that the Auger process is determined by nucleonic
particle-hole fluctuations of the nuclear mean-field absorbing the
energy-momentum transfer from the $\Lambda$ transition. This
structure becomes even more obvious by expressing the matrix
elements in terms of the appropriate non-diagonal elements of
$\Lambda$ and nucleon one-body density matrices in momentum space:
\begin{equation}
  \label{eq:mat_el}
  \left< 0\right|
  a_{\Lambda_\beta}a_{n_\beta}V a^\dag_{n^{-1}_\beta}a^\dag_{\Lambda_\alpha}
  \left| 0 \right>
  =\int d^4q\; \rho_{\Lambda'\Lambda}(q)V(q)\rho_{n^{-1}n}(q)
  \quad .
\end{equation}
Here, the transition densities $\rho_{ij}$ are given by:
\begin{eqnarray}
  \label{eq:trans_ff}
  \rho_{ij}(q) &\equiv& \int d^4x
  e^{iqx}\overline{\psi}_i(x)\hat\Gamma\psi_j(x) \nonumber\\
  &=& \delta\left((q^0 - (E_i -E_j)\right) \int d\Omega\;dr\;r^2
  \left(\sum_{\mu,\lambda}(-)^\lambda Y^*_{\lambda\mu}(\hat q)
    Y_{\lambda\mu}(\hat r)
    j_\lambda(qr)\right)\overline{\psi}_i(r,\hat r)\hat\Gamma
  \psi_j(r,\hat r)
\end{eqnarray}
$\hat\Gamma$ is either $\hat 1$ or $\gamma^\mu$ for $\sigma$- and
$\omega$-exchange, respectively. $E_{i,j}$ are the single particle energies
of the states $\psi_{i,j}$. In the second line the spatial
part of the plane wave is expanded into partial waves. A more
detailed description of the evaluation of the matrix elements is
given in appendix~\ref{asec:Matrixel}. The wave functions and
single particle energies $E_{i,j}$ are taken from a DDRH calculation which
is described in section~\ref{ssec:model}.

\subsection{\label{ssec:model}The DDRH Model}
%
The density dependent relativistic hadron field theory (DDRH)
\cite{Fuchs:1995as,Hofmann:2001vz} is a relativistic Lagrangian
field theory with baryons and mesons interacting by density
dependent coupling functionals. The meson-baryon interaction part
of the DDRH-Lagrangian used here is chosen as in
\cite{Fuchs:1995as}:
\begin{eqnarray}
  \label{eq:L_int}
  \mathcal{L}_{int} &=& \overline{\Psi}
  \Gamma_\sigma(\hat\rho) \Psi \sigma
  - \overline{\Psi} \Gamma_\omega(\hat\rho) \gamma_\mu \Psi \omega^\mu
  \nonumber\\
  &&- \frac{1}{2}\overline{\Psi}
  \vec\Gamma_\rho \gamma_\mu \Psi
  \vec\rho^\mu
  - e \overline{\Psi}_F \hat{Q} \gamma_\mu \Psi_F A^\mu
\end{eqnarray}
including isoscalar interactions from the scalar $\sigma$ and the
vector $\omega$ meson and isovector vector interactions given by
the $\rho$-meson. The last term accounts for the
electromagnetic interaction where $\hat{Q}$ is the charge
operator. The density dependence of the baryon-meson vertices is
determined self-consistently in mean-field approximation from
Dirac-Brueckner $(DB)$ self-energies by first solving the
Bethe-Salpeter equation for the in-medium two-particle scattering
amplitude and then extracting the medium modified vertices by
expanding the DB Hartree-Fock self-energies in terms of the
meson-exchange self-energies obtained from equation~(\ref{eq:L_int})
\cite{Fuchs:1995as,deJong98}. Medium modifications contribute at
various levels, e.g. the dressing of the baryon propagators by
self-energies and the Pauli-blocking of intermediate states. A
Lorentz invariant and thermodynamically consistent field theory is
retained by expressing the density dependent vertices in terms of
functionals of Lorentz scalar combinations $\hat\rho_i$ of the
baryon field operators $\psi$ \cite{Fuchs:1995as}. The description
of infinite nuclear matter and pure isospin nuclei within DDRH
theory has been shown to be very good using the Bonn A and the
Groningen NN-potentials \cite{Fuchs:1995as,Hofmann:2001vz}.

In \cite{Keil:2000hk} the extension of the DDRH theory to the
strangeness sector and applications to hypernuclear structure
calculations were discussed. Since DB calculations describing the
in-medium interactions of the complete baryon octet are not
available we follow here the approximation scheme discussed also
in \cite{Keil:2000hk}. The essential step is obtained from the
diagrammatic analysis of DB hyperon self-energies, showing that
the coupling functionals of nucleons and hyperons are related in
leading order by a simple scaling law given by the ratio of the
free space meson-nucleon and meson-hyperon  Born-term coupling
constants $g_{N\alpha}$ and $g_{Y\alpha}$, respectively \cite{Keil:2000hk}:
\begin{equation}
\Gamma_{Y\alpha}(\rho_Y) \approx 
\frac{g_{Y\alpha}}{g_{N\alpha}}\Gamma_{N\alpha}(\rho_Y)
\equiv R_{Y\alpha}\Gamma_{N\alpha}(\rho_Y).
\end{equation}
The scalar scaling
factor $R_{\Lambda\sigma}$ was taken from the J\"ulich-model
($R_{\Lambda\sigma}=0.49$)
\cite{Haidenbauer:1998kk}, including explicitly $\pi\pi$ and
$K\overline{K}$ dynamical correlations 
in the $0^+$ scalar-isoscalar meson channel. The scaling factor
for the $\Lambda\Lambda\omega$ vector vertex is obtained
phenomenologically from a least-square fit to $\Lambda$ separation
energies, resulting in $R_{\Lambda\omega} = 0.553$
\cite{Keil:2000hk}. However, in ref.\cite{Keil:2000hk} it was
noticed that the binding energies in the light nuclei below oxygen
seem to behave differently from the systematics obtained from
$A\geq 40$. The $\Lambda$ separation energies in the light nuclei
are described with the same accuracy only if the repulsion from
the $\omega$ meson is reduced by about 5\% to $R_{\Lambda\omega} =
0.542$. DDRH results for single $\Lambda$ hypernuclei are compared
to the presently available set of world data in
\cite{Keil:2000hk}. An agreement on the percent level is
obtained, being at least of the same quality as those of purely
phenomenological mean-field models.


\section{\label{sec:results}Results}

In this work we consider the hypernucleus $^{209}_\Lambda$Pb as
representative for the heavy mass hypernuclei (it is one of the nuclei that
are going to be studied in the JLAB experiment). The wave functions
and single particle energies used for the evaluation of the matrix
elements are calculated self-consistently by solving the DDRH
field equations in relativistic mean-field (RMF) approximation.
Details of the numerical approach and the model parameters are
given in \cite{Keil:2000hk}. 
For the present
application we need information on knocked-out, unbound neutron
states. For that purpose the single particle continuum was
discretized by enclosing the system in a huge box of size $R=150
fm$. Since the spacing of the discretrized continuum levels
behaves as $\sim \mathcal{O}(1/R^2)$ the use of such a large
quantization volume ensures quasi-continuous energy spectra for
the neutron scattering states, allowing to resolve single particle
resonances and other continuum structures resulting from the
calculations.

The continuum wave functions are calculated in the
self-consistently obtained ground state mean-field potentials.
Thus, final state interactions are taken into account on the level
of static mean-field self-energies. The approach assures
orthogonality of bound and unbound wave functions thus avoiding
the unphysical non-orthogonality contributions inherent to
phenomenological approaches.


Since the previous applications of the DDRH theory show that the
experimental $\Lambda$ spectra are reproduced especially accurate
in heavy nuclei our approach is well suited for the calculation of
the Auger neutron spectra from $^{209}_\Lambda$Pb. Numerical values of the
single particle energies for occupied neutron states and bound
$\Lambda$ states are displayed in
table~\ref{tab:single-particle-energies}.
\begin{table}[htbp]
  \begin{center}
    \begin{tabular}{l|r||l|r}
      $\Lambda$&E [MeV]&n&E [MeV] \\ \hline
      1s$_{1/2}$ & -27.16352 & 1g$_{9/2}$ & -27.67206\\
      1p$_{3/2}$ & -23.30941 & 1g$_{7/2}$ & -24.97048\\
      1p$_{1/2}$ & -23.08789 & 2d$_{5/2}$ & -21.91610\\
      1d$_{5/2}$ & -18.53900 & 2d$_{3/2}$ & -20.61122\\
      1d$_{3/2}$ & -17.98296 & 3s$_{1/2}$ & -19.86137\\
      2s$_{1/2}$ & -15.74262 & 1h$_{11/2}$ & -18.56635\\
      1f$_{7/2}$ & -13.15220 & 1h$_{9/2}$ & -14.83666\\
      1f$_{5/2}$ & -12.11529 & 2f$_{7/2}$ & -12.70419\\
      2p$_{3/2}$ & -9.64314  & 2f$_{5/2}$ & -10.93869\\
      2p$_{1/2}$ & -9.25274  & 3p$_{1/2}$ & -10.32673\\
      1g$_{9/2}$ & -7.40907  & 3p$_{3/2}$ & -9.64888\\
      1g$_{7/2}$ & -5.80963  & 1i$_{13/2}$ & -9.41419\\
      2d$_{5/2}$ &  -4.00386 &  \\
      2d$_{3/2}$ &  -3.42605 &  \\
      3s$_{1/2}$ &  -3.09504 &  \\
      1h$_{11/2}$ & -1.56587 &  \\

    \end{tabular}
    \caption{DDRH results for
      $\Lambda$ and neutron single particle energies in
      $^{209}_\Lambda$Pb entering
      into the Auger-calculations.  The standard set of interaction
      parameters was used \cite{Keil:2000hk}.}
    \label{tab:single-particle-energies}
  \end{center}
\end{table}

\begin{figure}[htbp]
  \begin{minipage}{.48\linewidth}
  \begin{center}
    \includegraphics[width=.95\linewidth]{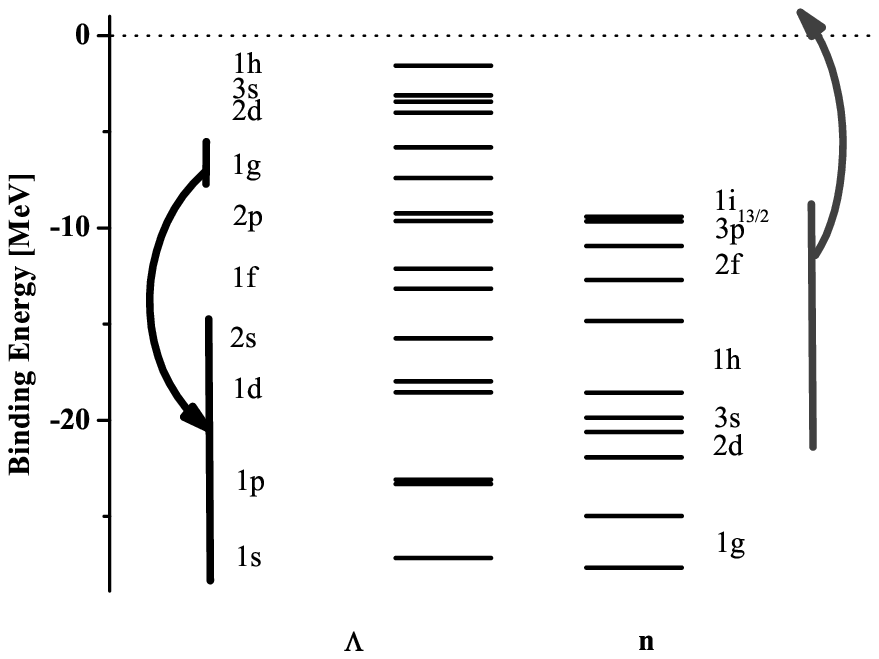}
    \caption{Level schemes of the $\Lambda$ hyperons and neutrons that are
      involved in the hypernuclear Auger effect. The levels are displayed in
      the physical scale as shown in table~\ref{tab:single-particle-energies}.
      The bars mark the single particle levels involved by
      the de-excitation of an initial 1g-shell $\Lambda$. The maximum
      energy, released when the $\Lambda$ drops down to the $1s$-orbit,
      allows to emit neutrons from the $1i_{13/2}$ valence orbit down to
      the $2d_{3/2}$ or $3s_{1/2}$ shells for the $1g_{7/2}$ and $1g_{9/2}$
      initial states, respectively.}
    \label{fig:auger-1g-levels}
  \end{center}
\end{minipage}\hfill
  \begin{minipage}{.48\linewidth}
  \begin{center}
    \includegraphics[width=.95\linewidth]{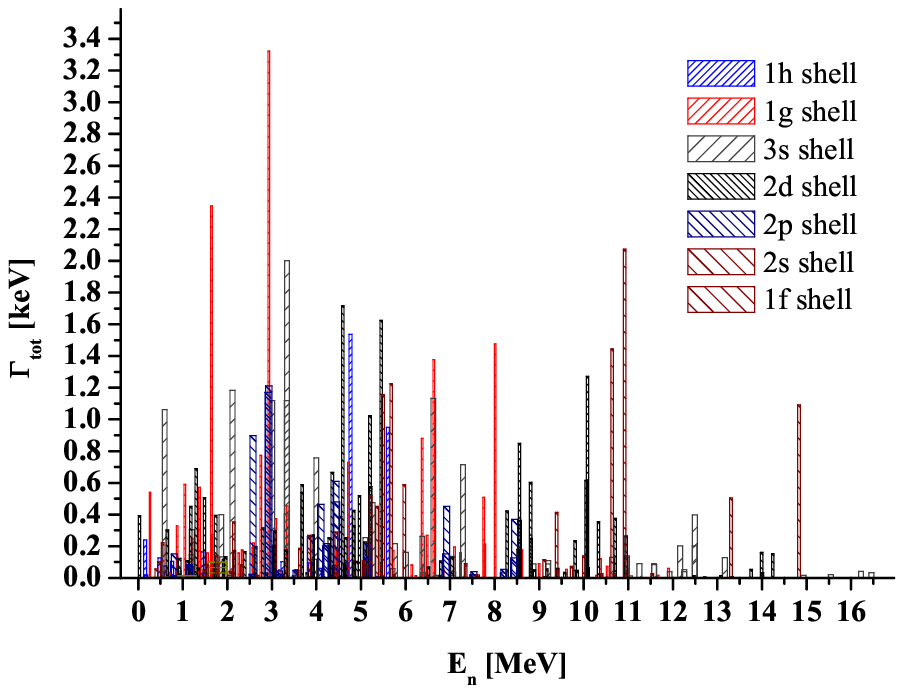}
    \caption{Full spectrum of the Auger transition strengths in
      $^{209}_\Lambda Pb$ including contributions by initial population
      of the $1h$, $1g$, $3s$, $2d$, $2p$, $2s$ and $1f$ $\Lambda$ orbitals
      and their subsequent de-excitation by neutron emission.}
    \label{fig:full-spec}
  \end{center}
\end{minipage}
\end{figure}
\begin{table}[htbp]
  \begin{center}
    \begin{tabular}{c|c||r|r}
      final $\Lambda$ & neutron-hole & $\Gamma_{1g_{9/2}} [10^{-2}keV]$ &
      $\Gamma_{1g_{7/2}} [10^{-2}keV]$ \\
      \hline
      $1s_{1/2}$ & $2d_{3/2}$ & -- & 3.2 \\
      & $3s_{1/2}$ & -- & 0.6 \\
      & $1h_{11/2}$ & 57.1 &  77.3 \\
      & $1h_{9/2}$ & 8.7 & 3.5 \\
      & $2f_{7/2}$ & 16.0 & 17.8 \\
      & $2f_{5/2}$ & 9.1 & 12.2 \\
      & $3p_{3/2}$ & 5.5 & 5.5 \\
      & $3p_{1/2}$ & 1.8 & 1.7 \\
      & $1i_{13/2}$ & 7.4 & 6.1 \\ &&& \\
      $1p_{3/2}$ & $1h_{9/2}$ & 24.8 & 22.1 \\
      & $2f_{7/2}$ & 45.3 & 73.0 \\
      & $2f_{5/2}$ &  8.7 & 26.9 \\
      & $3p_{3/2}$ & 17.3 & 19.6 \\
      & $3p_{1/2}$ & 8.6 & 21.5 \\
      & $1i_{13/2}$ & 137.8 & 147.6 \\ &&& \\
      $1p_{1/2}$ & $1h_{9/2}$ & 4.5 & 17.6 \\
      & $2f_{7/2}$ & 37.4 & 17.2 \\
      & $2f_{5/2}$ &  11.2 & 1.6 \\
      & $3p_{3/2}$ & 8.2 & 5.4 \\
      & $3p_{1/2}$ & 8.1 & 2.0 \\
      & $1i_{13/2}$ & 88.1 & 51.0 \\ &&& \\
      $1d_{5/2}$ & $2f_{5/2}$ & 54.1 & 234.6 \\
      & $3p_{3/2}$ & 32.9 & 15.7 \\
      & $3p_{1/2}$ & 15.9 & 332.2 \\
      & $1i_{13/2}$ & 9.4 & 5.3 \\ &&& \\
      $1d_{3/2}$ & $2f_{5/2}$ & -- & 59.0 \\
      & $3p_{3/2}$ & $<$0.01 & 27.5 \\
      & $3p_{1/2}$ & 0.5 & 17.2 \\
      & $1i_{13/2}$ & 1.2 & 19.2 \\ &&& \\
      $2s_{1/2}$ & $3p_{1/2}$ & -- & $<$0.01
    \end{tabular}
    \caption{Transitions contributing to the de-excitation of the
      $1g^\Lambda$-states. Transition widths for the $\Lambda$ initial
      states $1g_{9/2}$ and $1g_{7/2}$ are denoted by $\Gamma_{1g_{9/2}}$ and
      $\Gamma_{1g_{7/2}}$, respectively.}
    \label{tab:1g-trans}
  \end{center}
\end{table}

Because of the high level density and moderate separation energies
heavy hypernuclei are most suitable for the Auger effect. However,
at the same time these apparent advantages are, unfortunately, a
potential source of problems for experimental work. The huge
amount of combinatorial possibilities for transitions, illustrated
in table~\ref{tab:1g-trans} and indicated in
figure~\ref{fig:auger-1g-levels} for the case of initially
populating the $1g^\Lambda$-shell in $^{209}_\Lambda$Pb, leads to
Auger spectra of a rather complicated shape making in many cases
an unambiguous identification of transitions and assignment of
quantum numbers almost impossible.

The problem is apparent from figure~\ref{fig:full-spec} where the
complete neutron emission spectrum from the $^{209}_\Lambda Pb$
hypernucleus is displayed, summed over all energetically open
$\Lambda$ levels. Experimentally, spectra of a similar structure
have to be expected. In addition, state-dependent weighting
factors from the production vertex of the initial $\Lambda$ state
will be superimposed. From figure~\ref{fig:full-spec} and
figure~\ref{fig:auger-1g-levels} it is obvious that experiments
will be confronted with spectra of high complexity. Before
observables of physical interest can be accessed the data will
have to be analyzed in a more selective approach. From the
discussion it is obvious that much of the structure will be
produced by incoherent superpositions of contributions from the
variety of orbitals in which the $\Lambda$ was inititally
produced. By a precise energy tagging of the produced kaon and the
outgoing electron
(for the case of electromagnetic production of hypernuclei, as
will be the case at JLAB) it might be possible to determine the
initial $\Lambda$-state accurately enough and that only the
Auger-neutrons related to the de-excitation of that specific state
can be recorded selectively in a coincidence measurement.

\begin{figure}[htbp]
  \begin{center}
    \includegraphics[width=.4\linewidth]{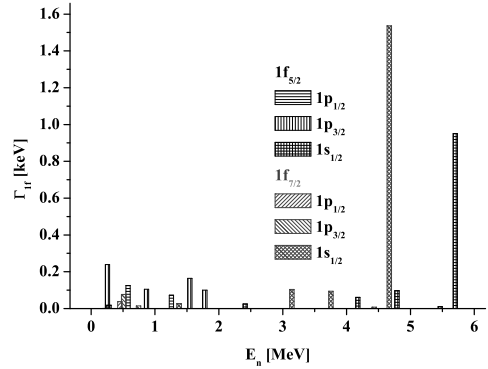}
    \caption{Decay widths of the Auger transition strengths in
      $^{209}_\Lambda Pb$ with an initial 1f $\Lambda$ state.}
    \label{fig:1f-spec}
  \end{center}
\end{figure}

In this context, it is of interest to consider the Auger
transitions on the limits of the energetically accessible range of
states. Assuming energetically sharp states, i.e. the $\Lambda$
and neutrons are in good quasiparticle configurations and damping
effects are negligible, the present calculation predicts that the
Auger process can only take place if the $\Lambda$ initially is
produced in the $2s^\Lambda$-orbit or above (see
table~\ref{tab:single-particle-energies}), because otherwise the
$\Lambda$ transition energies are less than the lowest neutron
separation energy. Next to the $2s^\Lambda$-orbit we find the
doublet of $1f^\Lambda$ states which is of interest because it
allows to observe the $\Lambda$ spin-orbit splitting, at least in
principle. Since the energy window available from populating the
$1f^\Lambda$-orbits is still rather narrow the resulting Auger
neutron spectrum is of a comparatively simple structure. In
figure~\ref{fig:1f-spec} results for the Auger spectra produced by
the $1f_{7/2}$ and $1f_{5/2}$ states are compared. It is seen that
in both cases only a small number of final states occurs. Even
more, only the transitions to the $1i_{13/2}$ neutron-hole final
state yield a significant strength. Thus it is even possible to
clearly obtain the spin-orbit splitting of the $1f^\Lambda$-shell.
The prominent strength of this doublet is also fairly model
independent. Considering the kinematically allowed phase space of
this transition only, one might expect that wave function effects
which are sensitive to the details of the interaction could
strongly influence the relative strengths between the emission of
the $1i$- or $3p$-shell neutrons, which are in energy almost
degenerate (see table~\ref{tab:single-particle-energies}).
Nevertheless, due to the high degeneracy of the $1i_{13/2}$
neutron-orbitals the multiplicity of these neutrons will be
greatly enhanced. Therefore one can expect to observe in the
de-excitation spectrum of a $1f^\Lambda$-state a clear line
doublet belonging to the two spin-orbit partners of the
$1f^\Lambda$ shell falling down to the 1s orbital and thereby
knocking out the $1i_{13/2}$ neutron.

\begin{figure}[htbp]
  \begin{center}
    \includegraphics[width=0.5\linewidth]{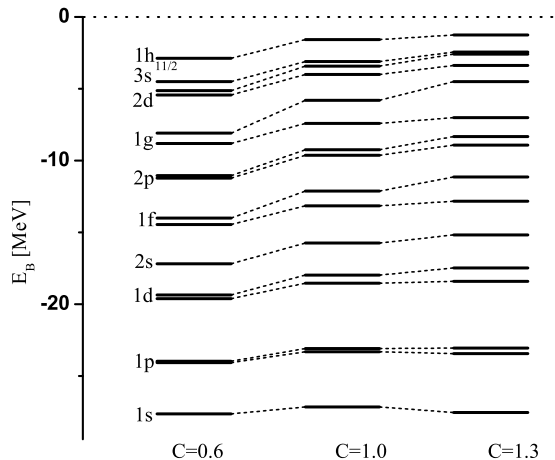}
    \caption{$\Lambda$ single particle energy spectra in $^{209}_\Lambda$Pb
      for the 
      three different sets of coupling constants 
    discussed in the text.}
    \label{fig:three-specs}
  \end{center}
\end{figure}
\begin{figure}[htbp]
  \begin{center}
    \includegraphics[width=.5\linewidth]{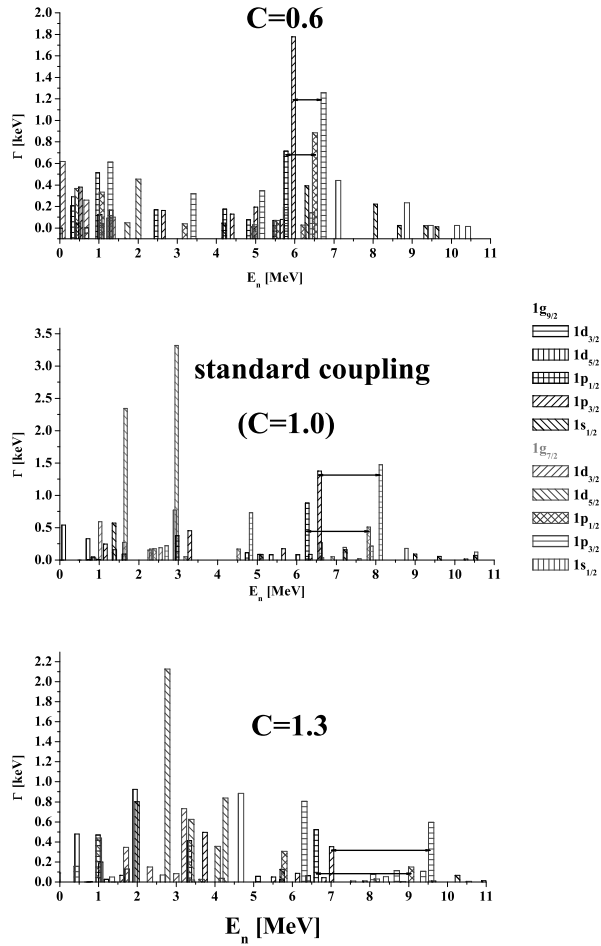}
    \caption{Comparison of the spectral Auger strength distributions
    produced by transitions from the $1g^\Lambda$ shell for interactions
      with different spin-orbit splittings. The Auger transition widths
      $\Gamma$ are shown as functions of 
      energy $E_n$ of the outgoing neutron.
      The upper and lower arrows indicate transitions from the 1g$_{9/2,7/2}$
      to the 
      1p$_{3/2}$ and 1p$_{1/2}$ final states, respectively.}
    \label{fig:coupling-comp}
  \end{center}
\end{figure}

The general case is, however, more ambiguous. The doublet
structure which one might expect as the signature of the $\Lambda$
spin-orbit splitting does not always show up in the spectral
distribution since in many cases the matrix elements depend
sensitively on binding energies and other wave function effects.
As a representative case, we study the influence of the spin-orbit
interaction strength on the Auger neutron spectra in more detail
for the $1g^\Lambda$ shell. This case is well suited, since the
spin-orbit splitting is sufficiently large and the spectrum offers
already some complexity. The possible transitions for this
configuration together with the corresponding transition rates are
shown in table~\ref{tab:1g-trans}.

The dependence of the $1g^\Lambda$-shell spectra on variations of
the spin-orbit strength of the $\Lambda$-nucleus potential is
investigated by changing the relative and the absolute coupling
strength of the $\sigma$ and the $\omega$ meson to the $\Lambda$,
keeping the overall single particle structure of the $\Lambda$
spectrum fixed. Numerically, this is realized by observing that in
our relativistic mean-field theory with scalar and vector
self-energies $U_\sigma$ and $U_\omega$, respectively, the leading
order non-relativistic Schroedinger-type central potential is
given by $U_0=U_\omega-U_\sigma$ and the strength of the
spin-orbit potential is determined by $U_{ls}=U_\omega+U_\sigma$.
Hence, we can relate spectral effects from variations of the
spin-orbit strength by a factor C to a scaling of the scalar and
vector self-energies $U_\sigma$ and $U_\omega$, respectively,
according to
\begin{eqnarray}
CU_{ls}&=&\beta U_\omega + \alpha U_\sigma \nonumber \\
DU_{0}&=&\beta U_\omega -\alpha U_\sigma  \quad .
\end{eqnarray}
We preserve the self-consistency between the nuclear mean-field
and the underlying baryon-baryon interactions by scaling the
$\Lambda\sigma$ and $\Lambda\omega$ vertices, equation~(\ref{eq:pot}), by
the same factors $\alpha$ and $\beta$, respectively.

Choosing $C$ as our external parameter and using always $D=1$ as a
constraint, the spin-orbit splittings of the $\Lambda$ levels can
be varied over wide ranges while keeping the changes in the
overall structure of the $\Lambda$ single particle spectrum on a
minimal level. The $\Lambda$ single particle spectra obtained for
$C=0.6,1.3$ (corresponding to
$(\alpha,\beta)=(0.6,0.5),(1.3,1,4)$, respectively,) are compared
to the results for $C=1$ in figure~\ref{fig:coupling-comp}. For
$C=0.6$ the spin-orbit splittings are reduced by about a factor of
2 and an increase of about the same size is found for $C=1.3$.
These (strong) variations will surely cover the full range of
uncertainties about the $\Lambda$ spin-orbit potential.

The spectral distributions of Auger transitions widths $\Gamma$,
obtained by putting the initial $\Lambda$ in $1g_{9/2,7/2}$ orbits
are displayed in figure~\ref{fig:coupling-comp}. Compared to the
standard case $C=1$ the spectral structures and transition
strengths are changed drastically when using $C=0.6$ and $C=1.3$.
As an overall feature the calculations show a concentration of
strength in a few states for $C=1$ and $C=1.3$ while a more
equilibrated distribution is found for $C=0.6$. The apparent
pile-up of strength at low neutron energies for the normal and
strong coupling cases are related to the variations in the
values of overlap matrix elements due to changes in binding
energies and single particle wave functions. Analyzing the
dynamical content of the response functions by calculating sum
rules for various moments of the excitation energy one finds a
disappointing small sensitivity on the spin-orbit interaction
strength. Hence, it is unlikely that spin-orbit effects will
contribute significantly and on an observable level to the total
spectral strength.

On the level of individual transitions there are, however, signals
for spin-orbit effects visible. In figure~\ref{fig:coupling-comp}
the transitions $1g^\Lambda_{9/2,7/2} \rightarrow
1p^\Lambda_{3/2}$ are indicated by the upper arrows while lower
arrows denote $1g^\Lambda_{9/2,7/2} \rightarrow 1p^\Lambda_{1/2}$.
The length of the arrows corresponds to the $1g^\Lambda_{9/2}
\leftrightarrow 1g^\Lambda_{7/2}$ energy splitting which obviously
depends directly on the spin-orbit interaction strength. In
addition, details of the spectral distributions, e.g. the
clustering of strength in certain energy regions, also depends on
the overlap of wave functions by which the transition matrix
elements and therefore the transition widths $\Gamma$ are
determined (see equations~(\ref{eq:differential-width2}) and
(\ref{eq:mat_el})).

It might seem somewhat disadvantageous to start out doing Auger experiments
with 
a system of the complexity of a heavy nucleus like lead. However, aiming at
resolving the $\Lambda$ spin-orbit structure in heavy hypernuclei, which is
still a controversial question as was mentioned in the introduction, one needs
to find hypernuclei that provide a core nucleus of very low spin. In the ideal
case this should be J$^\pi= 0^+$ in order to eliminate - or at least to
suppress 
- effects from $\Lambda$ angular momentum - core nucleus spin interactions
leading to an additional splitting superimposed on the pure $\Lambda$
spin-orbit 
splitting. In this respect the frequently used $^{89}$Y and $^{51}$V
hypernuclei 
are very unfavorable cases because the $^{88}$Y and $^{50}$V core nuclei carry
ground state spins of 4$^-$ and 6$^+$, respectively! The broadening of the
l-orbits observed in these nuclei [1] does not exhibit a proper $\vec{l} \cdot
\vec{s}$ systematics which might be related to contribution from residual
interactions involving the huge core spins [25]. Hyperisotopes matching the
requirements of having a 0$^+$ core ground state and appearing with at least
around 10\% natural abundancy would be $^{91}_{\Lambda}$Zr and
$^{119}_\Lambda$Sn. Lighter isotopes will gradually cease to exhibit Auger
de-excitations.

\section{\label{sec:conclusions}Summary and Conclusions}

The Auger-neutron transition rates for the de-excitation of single
$\Lambda$ hypernuclei on the example of $^{209}_\Lambda$Pb have
been calculated in relativistic DDRH theory. The initial state was
represented by a ground state $^{208}$Pb core with an attached
$\Lambda$ single particle state, the final state as a neutron
particle-hole excited $^{208}$Pb where the $\Lambda$ hyperon
occupies a lower lying single particle level and the neutron of
the particle-hole pair being unbound. Hence the process
corresponds to a decay of the initial single $\Lambda$
configuration into a $\Lambda$-neutron-particle-hole
configuration.

Due to the fact that the hypernuclear Auger effect appears mainly
in heavy and intermediate mass hypernuclei the spectral
distribution of the emitted neutrons is extremely complex. For
this reason a very detailed reconstruction of each event will have
to be done in possible experimental measurements. By focusing on
initial states with the $\Lambda$ in the $1f^\Lambda$-orbit our
calculations predict a clean doublet structure in the Auger
neutron spectrum from which the $1f^\Lambda$ spin-orbit splitting
can be directly read off, assuming sufficient energy resolution.
For a general $\Lambda$ orbit no definite signature of the
spin-orbit splitting will appear due to the huge amount of
transitions and the additional broadening of peaks when deeply
bound neutrons are involved. The effect of a finite width of the
neutron states, increasing with the distance from the
Fermi-surface has not been considered in our calculations.  For
the least bound neutrons it will have almost no effect, but going
deeper in binding energy the broadening of the states can no
longer be neglected. As described in the previous paragraphs the
neutron spectra in which the deeply bound neutrons are also
involved are already fairly complex. An additional broadening of
the neutron levels will thus wash out most of the spectral
structure there. However, in the case of the $1f^\Lambda$-shell
de-excitation only the valence neutrons close to the Fermi-level
are involved so that the spin-orbit splitting signal is going to
survive in realistic spectra.

For the spectroscopy in the intermediate and heavy mass region the
hypernuclear Auger effect provides a promising complementary tool
to $\gamma$ spectroscopy, although special care must be taken on
the choice of transitions and tagging the energy. For the
$^{209}_\Lambda$Pb hypernucleus the $1f^\Lambda$ shell is such a
well suited case. This technique might be the only way to achieve
high resolution information on the hypernuclear fine structure in
heavy nuclei since for $\gamma$ transitions it will be even harder
to assign the detected photons to specific transitions.

\section{\label{sec:Ackn}Acknowledgements}
We would like to thank Amour Margarian for exciting our interest
in the hypernuclear Auger effect and for useful discussions on
this topic. We would also like to thank Simona Briganti for useful
discussions on the general features of the atomic Auger effect.
This work was supported by the European Graduate School Gie\ss
en--Copenhagen and DFG-contract LE439/4b.

\appendix

\section{\label{asec:Matrixel}Matrix Elements}
For the evaluation of the matrix elements the transition form-factors,
eq.~(\ref{eq:trans_ff}), are evaluated in the spherical basis of
Dirac wave functions:
\begin{eqnarray}
  \label{eq:eval_trans-ff}
  \rho_{12}(q) &=& \int d^3x \; \overline{\psi}_1(\vec x)\hat\Gamma
  \psi_2(\vec x)\; e^{i\vec k \vec x} \nonumber\\
  &=& \sum_{\lambda\mu}(-)^\lambda Y^*_{\lambda\mu}(\hat q)
  \int d\Omega\;dr\;r^2\; j_\lambda(kr)Y_{\lambda\mu}(\hat r)
  \left(
    \begin{array}{c}
      g_1(r)\;\Omega_{j_1l_1m_1}(\theta,\phi) \\
      if_1(r)\;\Omega_{j_1\tilde l_1m_1}(\theta,\phi)
    \end{array}
  \right)^\dag
  \gamma^0\hat\Gamma
  \left(
    \begin{array}{c}
      g_2(r)\;\Omega_{j_2l_2m_2}(\theta,\phi) \\
      if_2(r)\;\Omega_{j_2\tilde l_2m_2}(\theta,\phi)
    \end{array}
  \right)
\end{eqnarray}
$\hat\Gamma$ is either the four by four unit matrix or
$\gamma_\mu$, depending whether the scalar or the vector potential
is evaluated. $\hat q$ and $\hat r$ denotes the unit vectors in
direction of $\vec q$ and $\vec r$, respectively. The spinors are
the usual total angular momentum eigenstates of the Dirac equation
with the generalized spin-angle spherical harmonics
$\Omega_{jlm}(\theta,\phi)$ obtained by coupling spin and orbital
angular momenta \cite{Bjorken-Drell,glaudemans}. The numerical
solution of the radial Dirac equation was discussed e.g. in
\cite{Fuchs:1995as}. The orbital angular momenta $l$ and $\tilde
l$ are determined by $j$ and the parity $\pi$:
\begin{equation}
  \label{eq:l-and-l-tilde}
    l=\left\{
      \begin{array}{l}
        j+1/2 \quad \text{ for}\; \pi=(-)^{j+1/2} \\
        j-1/2 \quad \text{ for}\; \pi=(-)^{j-1/2}
      \end{array}
    \right.
    \qquad
    \tilde l=\left\{
      \begin{array}{l}
        j-1/2 \quad \text{ for}\; \pi=(-)^{j+1/2} \\
        j+1/2 \quad \text{ for}\; \pi=(-)^{j-1/2}
      \end{array}
    \right.
\end{equation}
For the scalar vertex we get then
\begin{eqnarray}
  \label{eq:scalar-trans-ff}
  \rho^s_{12}(q) = \sum_{\lambda\mu}(-)^\lambda Y^*_{\lambda\mu}(\hat q)
  &&\left\{
    \left[ \int dr\;r^2\;g_1(r)g_2(r)j_\lambda(qr) \right]
    \left[ \int d\Omega\; \Omega^*_{j_1l_1m_1}Y_{\lambda\mu}
      \Omega_{j_2l_2m_2}\right]
  \right. \nonumber\\
  &&\left.
    - \left[ \int dr\;r^2\;f_1(r)f_2(r)j_\lambda(qr) \right]
    \left[ \int d\Omega\; \Omega^*_{j_1\tilde l_1m_1}Y_{\lambda\mu}
      \Omega_{j_2\tilde l_2m_2}\right]
  \right\}
\end{eqnarray}
The radial matrix element is evaluated numerically. The angular
integral can be performed analytically. By means of the
Wigner-Eckhardt theorem \cite{glaudemans} it can be expressed by
Clebsch-Gordan coefficients and reduced matrix elements
\begin{equation}
  \label{eq:reduced-mat-el-scalar}
  \int d\Omega\; \Omega^*_{jlm}Y_{\lambda\mu}\Omega_{j'l'm'}
  =\left<jlm \left| Y_{\lambda\mu}\right| j'l'm'\right>
  = (-)^{j-m}
  \left(
    \begin{array}{rrr}
      j&\lambda&j' \\ -m&\mu&m'
    \end{array}
  \right)
  \left<\left. l \frac{1}{2} j \right|\right|
  Y_\lambda \left|\left| l \frac{1}{2} j'\right>\right.
\end{equation}
For the vector transition form factor we have to evaluate
\begin{eqnarray}
  \label{eq:vector-trans-ff}
  \rho^{\nu}_{12}(q) = \sum_{\lambda\mu}(-)^\lambda Y^*_{\lambda\mu}(\hat q)
  &&\left\{
    \left[ \int dr\;r^2\;g_1(r)g_2(r)j_\lambda(qr) \right]
    \left[ \int d\Omega\; \Omega^*_{j_1l_1m_1}Y_{\lambda\mu}
      \Omega_{j_2l_2m_2}\right]
  \right. \nonumber\\
  &&
  + \left[ \int dr\;r^2\;f_1(r)f_2(r)j_\lambda(qr) \right]
  \left[ \int d\Omega\; \Omega^*_{j_1\tilde l_1m_1}Y_{\lambda\mu}
    \Omega_{j_2\tilde l_2m_2}\right], \nonumber\\
  &&{\bm i}
  \left[ \int dr\;r^2\;g_1(r)f_2(r)j_\lambda(qr) \right]
  \left[ \int d\Omega\; \Omega^*_{j_1l_1m_1}Y_{\lambda\mu}
    \vec\sigma\;\Omega_{j_2\tilde l_2m_2}\right]
  \nonumber\\
  &&\left.
    + \left[ \int dr\;r^2\;f_1(r)g_2(r)j_\lambda(qr) \right]
    \left[ \int d\Omega\; \Omega^*_{j_1\tilde l_1m_1}Y_{\lambda\mu}
      \vec\sigma\;\Omega_{j_2l_2m_2}\right]
  \right\}^\nu
\end{eqnarray}
The $\sigma$ are the usual Pauli matrices.
The reduced matrix element for the angular $\rho^0_{12}$ matrix element
is identical to equation~(\ref{eq:reduced-mat-el-scalar}), for the spatial
components it is given by
\begin{eqnarray}
  \label{eq:reduced-mat-el-vector}
  \left<jlm \right| Y_{\lambda\mu}\sigma_M \left| j'l'm' \right> &=&
  \sum_{IN} \left<\lambda\mu 1 M\right|\left. IN\right>
  \left< jlm\right|\left|Y_{\lambda\mu}\sigma_{1M}\right]_{IN}
  \left| j'l'm'\right> \nonumber\\
  &=& \sum_{IN}(-)^{j-m}  \left<\lambda\mu 1 M\right|\left. IN\right>
  \left(
    \begin{array}{rrrrrr}
      j&I&j' \\ m&N&m'
    \end{array}
  \right)
  \left<\left. l \frac{1}{2}j\right|\right|
  \left[Y_\lambda \sigma_1 \right]_I
  \left|\left| l'\frac{1}{2} j'\right>\right.\nonumber\\
  &=& \sum_{IN}(-)^{j-m}  \left<\lambda\mu 1 M\right|\left. IN\right>
  \left(
    \begin{array}{rrrrrr}
      j&I&j' \\ m&N&m'
    \end{array}
  \right)
  \left\{
    \begin{array}{rrrrrrrrr}
      l&\frac{1}{2}&j \\ l&\frac{1}{2}&j' \\ \lambda&1&I
    \end{array}
  \right\}
  \left<\left. l \right|\right| Y_\lambda \left|\left| l' \right>\right.
  \left<\left. \frac{1}{2}\right|\right|\sigma\left|\left|\frac{1}{2}\right>\right.
\end{eqnarray}
Note that the Pauli matrices $\sigma_M$, $M=0,\pm 1$, are used
here in the spherical basis \cite{glaudemans}. Explicit expessions
for the reduced matrix elements are found e.g. in ref.
\cite{glaudemans}.

%


\begin{thebibliography}{10}
\bibitem{Hotchi:2001rx}
H.~Hotchi {\it et al.},
Phys.\ Rev.\ C {\bf 64}, 044302 (2001).
\bibitem{nagae-hyp2000}
  T.~Nagae, Nucl.Phys.A {\bf 691}, 76c (2001) {\it and}
  H.~Hotchi, KEK Report 2000-3, April 2000
\bibitem{may-c13}
  M.~May {\it et al.}, Phys. Rev. Lett. {\bf 78}, 4343 (1997)
\bibitem{Hasegawa:fj}
  T.~Hasegawa {\it et al.},
  Phys.\ Rev.\ C {\bf 53}, 1210 (1996).
\bibitem{Ajimura:1994jb}
  S.~Ajimura {\it et al.},
  Nucl.\ Phys.\ A {\bf 585}, 173C (1995).
\bibitem{Pile:cf}
  P.~H.~Pile {\it et al.},
  Phys.\ Rev.\ Lett.\  {\bf 66}, 2585 (1991).
\bibitem{Hjorth-Jensen:1996yj}
M.~Hjorth-Jensen, A.~Polls, A.~Ramos and H.~Muther,
Nucl.\ Phys.\ A {\bf 605}, 458 (1996)
[arXiv:nucl-th/9604028].
\bibitem{Vidana:1998ed}
I.~Vidana, A.~Polls, A.~Ramos and M.~Hjorth-Jensen,
Nucl.\ Phys.\ A {\bf 644}, 201 (1998)
[arXiv:nucl-th/9805032].
\bibitem{Rufa:uf}
M.~Rufa, H.~Stocker, J.~A.~Maruhn, W.~Greiner and P.~G.~Reinhard,
J.\ Phys.\ G {\bf 13}, L143 (1987).
\bibitem{Rufa:zm}
M.~Rufa, J.~Schaffner, J.~Maruhn, H.~Stoecker, W.~Greiner and P.~G.~Reinhard,
Phys.\ Rev.\ C {\bf 42}, 2469 (1990).
\bibitem{Glendenning:1992du}
N.~K.~Glendenning, D.~Von-Eiff, M.~Haft, H.~Lenske and M.~K.~Weigel,
Phys.\ Rev.\ C {\bf 48}, 889 (1993)
[arXiv:nucl-th/9211012].
\bibitem{Millener:2000th}
D.~J.~Millener,
Nucl.\ Phys.\ A {\bf 691}, 93 (2001)
[arXiv:nucl-th/0103017].
\bibitem{Keil:2001rm}
  C.~Keil and H.~Lenske,
  APS Conf.Proc {\bf 603} (Mesons and Light Nuclei 2001), 433 (2001)
  [arXiv:nucl-th/0107068].
\bibitem{Keil:2000hk}
  C.~M.~Keil, F.~Hofmann and H.~Lenske,
  Phys.\ Rev.\ C {\bf 61}, 064309 (2000)
  [arXiv:nucl-th/9911014].
\bibitem{Kohri:2002nc}
  H.~Kohri {\it et al.}  [AGS-E929 Collaboration],
  Phys.\ Rev.\ C {\bf 65}, 034607 (2002)
  [arXiv:nucl-ex/0110007].
\bibitem{Likar:1986jj}
  A.~Likar, M.~Rosina and B.~Povh,
  Z.\ Phys.\ A {\bf 324}, 35 (1986).
\bibitem{Margaryan}
  A.~Margaryan, L.~Tang, S.~Majewski, O.~Hashimoto, V.~Likachev,
  {\em Auger Neutron Spectroscopy of Nuclear Matter at CEBAF},
  Letter of intent to JLAB PAC 18, LOI-00-101, 2000
\bibitem{pdg} see e.g. Particle Data Book, sect.~34, p.~211,
  Rev.~Part.~Phys., Europhys.J. {\bf C} 15 (2000) 1
\bibitem{glaudemans}
  P.J.~Brussaard, P.W.M.~Glaudemans,
  {\em Shell-model applications in nuclear spectroscopy},
  North-Holland, 1977.
\bibitem{Fuchs:1995as}
  C.~Fuchs, H.~Lenske and H.~H.~Wolter,
  Phys.\ Rev.\ C {\bf 52}, 3043 (1995)
  [arXiv:nucl-th/9507044],
  H.~Lenske and C.~Fuchs, Phys.Lett.B {\bf 345} 355 (1995).
\bibitem{Hofmann:2001vz}
  F.~Hofmann, C.~M.~Keil and H.~Lenske,
  Phys.\ Rev.\ C {\bf 64}, 034314 (2001)
  [arXiv:nucl-th/0007050].
\bibitem{deJong98}F. de Jong, H. Lenske, Phys.\ Rev.\ C {\bf 57},
3099 (1997)
\bibitem{Haidenbauer:1998kk}
  J.~Haidenbauer, W.~Melnitchouk and J.~Speth,
  arXiv:nucl-th/9805014.
\bibitem{Bjorken-Drell}J.D. Bjorken, S.D. Grell, Relativistic
Quantum Mechanics, McGraw-Hill, New York, 1964.
\bibitem{Keil:HotchiInterpretation} This has been found analysing the assumed
  spin-orbit splittings of \cite{Hotchi:2001rx} within DDRH and phen. RMF. Publication in preparation.

\end{thebibliography}
\end{document}